\documentclass[a4paper,11pt,onecolumn,nofootinbib,preprint]{article}
\usepackage{amssymb,amsmath,epsfig,subfig} 
\usepackage{times}
\usepackage{cite}
\usepackage{hyperref}
\usepackage{graphicx}
\usepackage{url}
\usepackage{slashed,cancel,ulem}
\usepackage{color}
\def\baselinestretch{1.4}
\begin{document}
 \topmargin -0.1in
\headsep 30pt
\footskip 40pt
\oddsidemargin 12pt
\evensidemargin -16pt
\textheight 8.5in
\textwidth 6.5in
\parindent 20pt
 
\def\baselinestretch{1.5}
\newcommand{\newc}{\newcommand}
\def\preprint{{preprint}}
\def\Ord{\lower .7ex\hbox{$\;\stackrel{\textstyle <}{\sim}\;$}}
\def\OOrd{\lower .7ex\hbox{$\;\stackrel{\textstyle >}{\sim}\;$}}
\def\cO#1{{\cal{O}}\left(#1\right)}
\newc{\order}{{\cal O}}
\def\lag             {{\cal L}}
\def\Lag             {{\cal L}}
\def\lum             {{\cal L}}
\def\R               {{\cal R}}
\def\Rsq             {{\cal R}^{\sq}}
\def\Rst             {{\cal R}^{\st}}
\def\Rsb             {{\cal R}^{\sb}}
\def\M               {{\cal M}}
\def\Oas             {{\cal O}(\alpha_{s})}
\def\Vcal            {{\cal V}}
\def\Wcal            {{\cal W}}
\newc{\be}{\begin{equation}}
\newc{\ee}{\end{equation}}
\newc{\br}{\begin{eqnarray}}
\newc{\er}{\end{eqnarray}}
\newc{\ba}{\begin{array}}
\newc{\ea}{\end{array}}
\newc{\bi}{\begin{itemize}}
\newc{\ei}{\end{itemize}}
\newc{\bn}{\begin{enumerate}}
\newc{\en}{\end{enumerate}}
\newc{\bc}{\begin{center}}
\newc{\ec}{\end{center}}
\newc{\ul}{\underline}
\newc{\ra}{\rightarrow}
\newc{\lra}{\longrightarrow}
\newc{\wt}{\widetilde}
\newc{\til}{\tilde}
\def\kr              {^{\dagger}}
\newc{\wh}{\widehat}
\newc{\ti}{\times}
\newc{\Dir}{\kern -6.4pt\Big{/}}
\newc{\Dirin}{\kern -10.4pt\Big{/}\kern 4.4pt}
\newc{\DDir}{\kern -10.6pt\Big{/}}
\newc{\DGir}{\kern -6.0pt\Big{/}}
\newc{\sig}{\sigma}
\newc{\sigmalstop}{\sig_{\lstoppair}}
\newc{\Sig}{\Sigma}  
\newc{\del}{\delta}
\newc{\Del}{\Delta}
\newc{\lam}{\lambda}
\newc{\Lam}{\Lambda}
\newc{\gam}{\gamma}
\newc{\Gam}{\Gamma}
\newc{\eps}{\epsilon}
\newc{\Eps}{\Epsilon}
\newc{\kap}{\kappa}
\newc{\Kap}{\Kappa}
\newc{\modulus}[1]{\left| #1 \right|}
\newc{\eq}[1]{(\ref{eq:#1})}
\newc{\eqs}[2]{(\ref{eq:#1},\ref{eq:#2})}
\newc{\etal}{{\it et al.}\ }
\newc{\ibid}{{\it ibid}.}
\newc{\ibidem}{{\it ibidem}.}
\newc{\eg}{{\it e.g.}\ }
\newc{\ie}{{\it i.e.}\ }
\def \viz{\emph{viz.}}
\def \etc{\emph{etc. }}
\newc{\nonum}{\nonumber}
\newc{\lab}[1]{\label{eq:#1}}
\newc{\dpr}[2]{({#1}\cdot{#2})}
\newc{\lt}{\stackrel{<}}
\newc{\gt}{\stackrel{>}}
\newc{\lsimeq}{\stackrel{<}{\sim}}
\newc{\gsimeq}{\stackrel{>}{\sim}}
\def\lsim{\buildrel{\scriptscriptstyle <}\over{\scriptscriptstyle\sim}}
\def\gsim{\buildrel{\scriptscriptstyle >}\over{\scriptscriptstyle\sim}}
\def\lapp{\mathrel{\rlap{\raise.5ex\hbox{$<$}}
                    {\lower.5ex\hbox{$\sim$}}}}
\def\gapp{\mathrel{\rlap{\raise.5ex\hbox{$>$}}
                    {\lower.5ex\hbox{$\sim$}}}}
\newc{\half}{\frac{1}{2}}
\newcommand {\nnc}        {{\overline{\mathrm N}_{95}}}
\newcommand {\dm}         {\Delta m}
\newcommand {\dM}         {\Delta M}
\def\bra{\langle}
\def\ket{\rangle}
\def\cO#1{{\cal{O}}\left(#1\right)}
\def \DM{{\Delta{m}}}
\newc{\bQ}{\ol{Q}}
\newc{\dota}{\dot{\alpha }}
\newc{\dotb}{\dot{\beta }}
\newc{\dotd}{\dot{\delta }}
\newc{\nindnt}{\noindent}

\newcommand{\medf}[2] {{\footnotesize{\frac{#1}{#2}} }}
\newcommand{\smaf}[2] {{\textstyle \frac{#1}{#2} }}
\def\onesq            {{\textstyle \frac{1}{\sqrt{2}} }}
\def\onehf            {{\textstyle \frac{1}{2} }}
\def\oneth            {{\textstyle \frac{1}{3} }}
\def\twoth            {{\textstyle \frac{2}{3} }}
\def\onefo            {{\textstyle \frac{1}{4} }}
\def\forth            {{\textstyle \frac{4}{3} }}

\newc{\matth}{\mathsurround=0pt}
\def\ML{\ifmmode{{\mathaccent"7E M}_L}
             \else{${\mathaccent"7E M}_L$}\fi}
\def\MR{\ifmmode{{\mathaccent"7E M}_R}
             \else{${\mathaccent"7E M}_R$}\fi}
\newcommand{\s}{\\ \vspace*{-3mm} }

\def \ud { {1 \over 2} }
\def \ut { {1 \over 3} }
\def \td { {3 \over 2} }
\newc{\mr}{\mathrm}
\def\dh {\partial }
\def \cs { cross-section }
\def \css { cross-sections }
\def \cm { centre of mass }
\def \cms { centre of mass energy }
\def \cc { coupling constant }
\def \ccs {coupling constants }
\def \gc {gauge coupling }
\def \gcc {gauge coupling constant }
\def \gccs {gauge coupling constants }
\def \yc {Yukawa coupling }
\def \ycc {Yukawa coupling constant }
\def \pp {{parameter }}
\def \pps {{parameters }} 
\def \ps {parameter space }
\def \pss {parameter spaces }
\def \vv {vice versa }

\newc{\siminf}{\mbox{$_{\sim}$ {\small {\hspace{-1.em}{$<$}}}    }}
\newc{\simsup}{\mbox{$_{\sim}$ {\small {\hspace{-1.em}{$>$}}}    }}


\newc {\Zboson}{{\mathrm Z}^{0}}
\newc{\thetaw}{\theta_W}
\newc{\mbot}{{m_b}}
\newc{\mtop}{{m_t}}
\newc{\sm}{${\cal {SM}}$}
\newc{\as}{\alpha_s}
\newc{\aem}{\alpha_{em}}
\def \PI{{\pi^{\pm}}}
\newc{\ppbar}{\mbox{$p\ol{p}$}}
\newc{\bbbar}{\mbox{$b\ol{b}$}}
\newc{\ccbar}{\mbox{$c\ol{c}$}}
\newc{\ttbar}{\mbox{$t\ol{t}$}}
\newc{\eebar}{\mbox{$e\ol{e}$}}
\newc{\zzero}{\mbox{$Z^0$}}
\def \gamz{\Gam_Z}
\newc{\wplus}{\mbox{$W^+$}}
\newc{\wminus}{\mbox{$W^-$}}
\newc{\ellp}{\ell^+}
\newc{\ellm}{\ell^-}
\newc{\elp}{\mbox{$e^+$}}
\newc{\elm}{\mbox{$e^-$}}
\newc{\elpm}{\mbox{$e^{\pm}$}}
\newc{\qbar}     {\mbox{$\ol{q}$}}
\def \ewgroup{SU(2)_L \otimes U(1)_Y}
\def \smgroup{SU(3)_C \otimes SU(2)_L \otimes U(1)_Y}
\def \smcolorem{SU(3)_C \otimes U(1)_{em}}

\def \SSM  {Supersymmetric Standard Model}
\def \poincare{Poincare$\acute{e}$}
\def \superspace{\emph{superspace}}
\def \sfs{\emph{superfields}}
\def \superpot{\emph{superpotential}}
\def \csf{\emph{chiral superfield}}
\def \csfs{\emph{chiral superfields}}
\def \vsf{\emph{vector superfield }}
\def \vsfs{\emph{vector superfields}}
\newc{\Ebar}{{\bar E}}
\newc{\Dbar}{{\bar D}}
\newc{\Ubar}{{\bar U}}
\newc{\susy}{{{SUSY}}}
\newc{\msusy}{{{M_{SUSY}}}}

\def\photino{\ifmmode{\mathaccent"7E \gam}\else{$\mathaccent"7E \gam$}\fi}
\def\taugluino{\ifmmode{\tau_{\mathaccent"7E g}}
             \else{$\tau_{\mathaccent"7E g}$}\fi}
\def\mphotino{\ifmmode{m_{\mathaccent"7E \gam}}
             \else{$m_{\mathaccent"7E \gam}$}\fi}
\newc{\gl}   {\mbox{$\wt{g}$}}
\newc{\mgl}  {\mbox{$m_{\gl}$}}
\def \charginopm{{\wt\chi}^{\pm}}
\def \mcharginopm{m_{\charginopm}}
\def \mchpmmin {\mcharginopm^{min}}
\def \chonep {{\wt\chi_1^+}}
\def \chone {{\wt\chi_1}}
\def \ch2p {{\wt\chi_2^+}}
\def \chonem {{\wt\chi_1^-}}
\def \ch2m {{\wt\chi_2^-}}
\def \chplus {{\wt\chi^+}}
\def \chminus {{\wt\chi^-}}
\def \chonip{{\wt\chi_i}^{+}}
\def \chonim{{\wt\chi_i}^{-}}
\def \chonipm{{\wt\chi_i}^{\pm}}
\def \chonjp{{\wt\chi_j}^{+}}
\def \chonjm{{\wt\chi_j}^{-}}
\def \chonjpm{{\wt\chi_j}^{\pm}}
\def \chonepm{{\wt\chi_1}^{\pm}}
\def \chonemp{{\wt\chi_1}^{\mp}}
\def \mchonepm{m_{\chonepm}}
\def \mchonemp{m_{\chonemp}}
\def \chtwopm{{\wt\chi_2}^{\pm}}
\def \mchtwopm{m_{\chtwopm}}
\newc{\dmchi}{\Delta m_{\wt\chi}}


\def \vlsp{\emph{VLSP}}
\def \lspi{\wt\chi_i^0}
\def \mlspi{m_{\lspi}}
\def \lspj{\wt\chi_j^0}
\def \mlspj{m_{\lspj}}
\def \lspone{\wt\chi_1^0}
\def \mlspone{m_{\lspone}}
\def \lsptwo{\wt\chi_2^0}
\def \mlsptwo{m_{\lsptwo}}
\def \lspthree{\wt\chi_3^0}
\def \mlspthree{m_{\lspthree}}
\def \lspfour{\wt\chi_4^0}
\def \mlspfour{m_{\lspfour}}


\newc{\sele}{\wt{\mathrm e}}
\newc{\sell}{\wt{\ell}}
\def \msell{m_{\sell}}
\def \slepone{\wt\ell_1}
\def \mslepone{m_{\slepone}}
\def \smuone{\wt\mu_1}
\def \msmuone{m_{\smuone}}
\def \stauone{\wt\tau_1}
\def \stauonepm{{\wt\tau_1}^\pm}
\def \mstauone{m_{\stauone}}
\def \snu{\wt{\nu}}
\def \snutau{\wt{\nu}_{\tau}}
\def \nutau{{\nu}_{\tau}}                  
\def \msnu{m_{\snu}}
\def \msnumu{m_{\snu_{\mu}}}
\def \barsnu{\wt{\bar{\nu}}}
\def \barsnul{\barsnu_{\ell}}
\def \snul{\snu_{\ell}}
\def \mbarsnu{m_{\barsnu}}
\newc{\snue}     {\mbox{$ \wt{\nu_e}$}}
\newc{\smu}{\wt{\mu}}
\newc{\stau}{\wt{\tau}}
\newc {\nuL} {\wt{\nu}_L}
\newc {\nuR} {\wt{\nu}_R}
\newc {\snub} {\bar{\wt{\nu}}}
\newc {\eL} {\wt{e}_L}
\newc {\eR} {\wt{e}_R}
\def \slepl{\wt{l}_L}
\def \mslepl{m_{\slepl}}
\def \slepr{\wt{l}_R}
\def \mslepr{m_{\slepr}}
\def \stau{\wt\tau}
\def \mstau{m_{\stau}}
\def \slepton{\wt\ell}
\def \mslepton{m_{\slepton}}
\def \mlhiggs{m_{h^0}}

\def \xr{X_{r}}

\def \sfer{\wt{f}}
\def \msfer{m_{\sfer}}
\def \sq{\wt{q}}
\def \msq{m_{\sq}}
\def \msquleft{m_{\tilde{u_L}}}
\def \msqurht{m_{\tilde{u_R}}}
\def \sql{\wt{q}_L}
\def \msql{m_{\sql}}
\def \sqr{\wt{q}_R}
\def \msqr{m_{\sqr}}
\newc{\msqot}  {\mbox{$m_(\sq_{1,2} )$}}
\newc{\sqbar}    {\mbox{$\bar{\wt{q}}$}}
\newc{\ssb}      {\mbox{$\squark\ol{\squark}$}}
\newc {\qL} {\wt{q}_L}
\newc {\qR} {\wt{q}_R}
\newc {\uL} {\wt{u}_L}
\newc {\uR} {\wt{u}_R}
\def \ul{\wt{u}_L}
\def \mul{m_{\ul}}
\newc {\dL} {\wt{d}_L}
\newc {\dR} {\wt{d}_R}
\newc {\cL} {\wt{c}_L}
\newc {\cR} {\wt{c}_R}
\newc {\sL} {\wt{s}_L}
\newc {\sR} {\wt{s}_R}
\newc {\tL} {\wt{t}_L}
\newc {\tR} {\wt{t}_R}
\newc {\stb} {\ol{\wt{t}}_1}
\newc {\sbot} {\wt{b}_1}
\newc {\msbot} {m_{\sbot}}
\newc {\sbotb} {\ol{\wt{b}}_1}
\newc {\bL} {\wt{b}_L}
\newc {\bR} {\wt{b}_R}
\def \mul{m_{\wt{u}_L}}
\def \mur{m_{\wt{u}_R}}
\def \mdl{m_{\wt{d}_L}}
\def \mdr{m_{\wt{d}_R}}
\def \mcl{m_{\wt{c}_L}}
\def \charml{\wt{c}_L}
\def \mcr{m_{\wt{c}_R}}
\newc{\csquark}  {\mbox{$\wt{c}$}}
\newc{\csquarkl} {\mbox{$\wt{c}_L$}}
\newc{\mcsl}     {\mbox{$m(\csquarkl)$}}
\def \msl{m_{\wt{s}_L}}
\def \msr{m_{\wt{s}_R}}
\def \mbl{m_{\wt{b}_L}}
\def \mbr{m_{\wt{b}_R}}
\def \mtl{m_{\wt{t}_L}}
\def \mtr{m_{\wt{t}_R}}
\def \st{\wt{t}}
\def \mst{m_{\st}}
\newc {\stopl}         {\wt{\mathrm{t}}_{\mathrm L}}
\newc {\stopr}         {\wt{\mathrm{t}}_{\mathrm R}}
\newc {\stoppair}      {\wt{\mathrm{t}}_{1}
\bar{\wt{\mathrm{t}}}_{1}}
\def \lstop{\wt{t}_{1}}
\def \lstopbar{\lstop^*}
\def \hstop{\wt{t}_{2}}
\def \hstopbar{\hstop^*}
\def \mlstop{m_{\lstop}}
\def \mhstop{m_{\hstop}}
\def \lstoppair{\lstop\lstop^*}
\def \hstoppair{\hstop\hstop^*}
\newc{\tsquark}  {\mbox{$\wt{t}$}}
\newc{\ttb}      {\mbox{$\tsquark\ol{\tsquark}$}}
\newc{\ttbone}   {\mbox{$\tsquark_1\ol{\tsquark}_1$}}
\def \tsq {top squark }
\def \tsqs {top squarks }
\def \tsql {ligtest top squark }
\def \tsqh {heaviest top squark }
\newc{\mix}{\theta_{\wt t}}
\newc{\cost}{\cos{\theta_{\wt t}}}
\newc{\sint}{\sin{\theta_{\wt t}}}
\newc{\costloop}{\cos{\theta_{\wt t_{loop}}}}
\def \lsbot{\wt{b}_{1}}
\def \lsbotbar{\lsbot^*}
\def \hsbot{\wt{b}_{2}}
\def \hsbotbar{\hsbot^*}
\def \mlsbot{m_{\lsbot}}
\def \mhsbot{m_{\hsbot}}
\def \lsbotpair{\lsbot\lsbot^*}
\def \hsbotpair{\hsbot\hsbot^*}
\newc{\mixsbot}{\theta_{\wt b}}

\def \mhone{m_{h_1}}
\def \hup{{H_u}}
\def \hdn{{H_d}}
\newc{\tb}{\tan\beta}
\newc{\cb}{\cot\beta}
\newc{\vev}[1]{{\left\langle #1\right\rangle}}

\def \abot{A_{b}}
\def \atop{A_{t}}
\def \atau{A_{\tau}}
\newc{\mhalf}{m_{1/2}}
\newc{\mzero} {\mbox{$m_0$}}
\newc{\azero} {\mbox{$A_0$}}

\newc{\lb}{\lam}
\newc{\lbp}{\lam^{\prime}}
\newc{\lbpp}{\lam^{\prime\prime}}
\newc{\rpv}{{\not \!\! R_p}}
\newc{\rpvm}{{\not  R_p}}
\newc{\rp}{R_{p}}
\newc{\rpmssm}{{RPC MSSM}}
\newc{\rpvmssm}{{RPV MSSM}}


\newc{\sbyb}{S/$\sqrt B$}
\newc{\pelp}{\mbox{$e^+$}}
\newc{\pelm}{\mbox{$e^-$}}
\newc{\pelpm}{\mbox{$e^{\pm}$}}
\newc{\epem}{\mbox{$e^+e^-$}}
\newc{\lplm}{\mbox{$\ell^+\ell^-$}}
\def \branch{\emph{BR}}
\def \branche{\branch(\lstop\ra be^{+}\nu_e \lspone)\ti \branch(\lstop^{*}\ra \bar{b}q\bar{q^{\prime}}\lspone)}
\def \branchmu{\branch(\lstop\ra b\mu^{+}\nu_{\mu} \lspone)\ti \branch(\lstop^{*}\ra \bar{b}q\bar{q^{\prime}}\lspone)}
\def\Ecm{\ifmmode{E_{\mathrm{cm}}}\else{$E_{\mathrm{cm}}$}\fi}
\newc{\rts}{\sqrt{s}}
\newc{\rtshat}{\sqrt{\hat s}}
\newc{\gev}{\,GeV}
\newc{\mev}{~{\rm MeV}}
\newc{\tev}  {\mbox{$\;{\rm TeV}$}}
\newc{\gevc} {\mbox{$\;{\rm GeV}/c$}}
\newc{\gevcc}{\mbox{$\;{\rm GeV}/c^2$}}
\newc{\intlum}{\mbox{${ \int {\cal L} \; dt}$}}
\newc{\call}{{\cal L}}
\def \met  {\mbox{${E\!\!\!\!/_T}$}}
\def \cpv  {\mbox{${CP\!\!\!\!/}$}}
\newc{\ptmiss}{/ \hskip-7pt p_T}
\def \eslash{\not \! E}
\def \etslash{\not \! E_T }
\def \ptslash{\not \! p_T }
\newc{\PT}{\mbox{$p_T$}}
\newc{\ET}{\mbox{$E_T$}}
\newc{\dedx}{\mbox{${\rm d}E/{\rm d}x$}}
\newc{\ifb}{\mbox{${\rm fb}^{-1}$}}
\newc{\ipb}{\mbox{${\rm pb}^{-1}$}}
\newc{\pb}{~{\rm pb}}
\newc{\fb}{~{\rm fb}}
\newc{\ycut}{y_{\mathrm{cut}}}
\newc{\chis}{\mbox{$\chi^{2}$}}
\def \hadron{\emph{hadron}}
\def \nlc{\emph{NLC }}
\def \lhc{\emph{LHC }}
\def \cdf{\emph{CDF }}
\def\dzero{\emptyset}
\def \tevatron{\emph{Tevatron }}
\def \lep{\emph{LEP }}
\def \jets{\emph{jets }}
\def \jet(s){\emph{jet(s) }}

\def\Crs{stroke [] 0 setdash exch hpt sub exch vpt add hpt2 vpt2 neg V currentpoint stroke 
hpt2 neg 0 R hpt2 vpt2 V stroke}
\def\loopdk{\lstop \ra c \lspone}
\def\brloopdk{\branch(\loopdk)}
\def\fourdk{\lstop \ra b \lspone  f \bar f'}
\def\brfourdk{\branch(\fourdk)}
\def\fourdklep{\lstop \ra b \lspone  \ell \nu_{\ell}}
\def\fourdkhad{\lstop \ra b \lspone  q \bar q'}
\def\brfourdklep{\branch(\fourdklep)}
\def\brfourdkhad{\branch(\fourdkhad)}
\def\twodk{\lstop \ra b \chonep}
\def\brtwodk{\branch(\twodk)}
\def\threedkslep{\lstop \ra b \wt{\ell} \nu_{\ell}}
\def\brthreedkslep{\branch(\threedkslep)}
\def\threedksnu{\lstop \ra b \wt{\nu_{\ell}} \ell }
\def\brthreedksnu{\branch(\threedksnu) }
\def\threedklsp{\lstop \ra b W \lspone }
\def\brthreedklsp{\\branch(\threedklsp) }
\def\topdk{t \ra \lstop \lspone}
\def\rpvdk{\lstop \ra e^+ d}
\def\brrpvdk{\branch(\rpvdk)}
\def\fonec{f_{11c}} 
\newc{\mpl}{M_{\rm Pl}}
\newc{\mgut}{M_{GUT}}
\newc{\mw}{M_{W}}
\newc{\mweak}{M_{weak}}
\newc{\mz}{M_{Z}}

\newc{\OPALColl}   {OPAL Collaboration}
\newc{\ALEPHColl}  {ALEPH Collaboration}
\newc{\DELPHIColl} {DELPHI Collaboration}
\newc{\XLColl}     {L3 Collaboration}
\newc{\JADEColl}   {JADE Collaboration}
\newc{\CDFColl}    {CDF Collaboration}
\newc{\DXColl}     {D0 Collaboration}
\newc{\HXColl}     {H1 Collaboration}
\newc{\ZEUSColl}   {ZEUS Collaboration}
\newc{\LEPColl}    {LEP Collaboration}
\newc{\ATLASColl}  {ATLAS Collaboration}
\newc{\CMSColl}    {CMS Collaboration}
\newc{\UAColl}    {UA Collaboration}
\newc{\KAMLANDColl}{KamLAND Collaboration}
\newc{\IMBColl}    {IMB Collaboration}
\newc{\KAMIOColl}  {Kamiokande Collaboration}
\newc{\SKAMIOColl} {Super-Kamiokande Collaboration}
\newc{\SUDANTColl} {Soudan-2 Collaboration}
\newc{\MACROColl}  {MACRO Collaboration}
\newc{\GALLEXColl} {GALLEX Collaboration}
\newc{\GNOColl}    {GNO Collaboration}
\newc{\SAGEColl}  {SAGE Collaboration}
\newc{\SNOColl}  {SNO Collaboration}
\newc{\CHOOZColl}  {CHOOZ Collaboration}
\newc{\PDGColl}  {Particle Data Group Collaboration}

\def\issue(#1,#2,#3){{\bf #1}, #2 (#3)}
\def\iss(#1,#2,#3){{\bf #1} (#3) #2}
\def\ASTR(#1,#2,#3){Astropart.\ Phys. \issue(#1,#2,#3)}
\def\AJ(#1,#2,#3){Astrophysical.\ Jour. \issue(#1,#2,#3)}
\def\AJS(#1,#2,#3){Astrophys.\ J.\ Suppl. \issue(#1,#2,#3)}
\def\APP(#1,#2,#3){Acta.\ Phys.\ Pol. \issue(#1,#2,#3)}
\def\JCAP(#1,#2,#3){Journal\ XX. \issue(#1,#2,#3)} 
\def\SC(#1,#2,#3){Science \issue(#1,#2,#3)}
\def\PRD(#1,#2,#3){Phys.\ Rev.\ D \issue(#1,#2,#3)}
\def\PR(#1,#2,#3){Phys.\ Rev.\ \issue(#1,#2,#3)} 
\def\PRC(#1,#2,#3){Phys.\ Rev.\ C \issue(#1,#2,#3)}
\def\NPB(#1,#2,#3){Nucl.\ Phys.\ B \issue(#1,#2,#3)}
\def\NPPS(#1,#2,#3){Nucl.\ Phys.\ Proc. \ Suppl \issue(#1,#2,#3)}
\def\NJP(#1,#2,#3){New.\ J.\ Phys. \issue(#1,#2,#3)}
\def\JP(#1,#2,#3){J.\ Phys.\issue(#1,#2,#3)}
\def\PL(#1,#2,#3){Phys.\ Lett. \issue(#1,#2,#3)}
\def\ZP(#1,#2,#3){Z.\ Phys. \issue(#1,#2,#3)}
\def\ZPC(#1,#2,#3){Z.\ Phys.\ C  \issue(#1,#2,#3)}
\def\PREP(#1,#2,#3){Phys.\ Rep. \issue(#1,#2,#3)}
\def\PRL(#1,#2,#3){Phys.\ Rev.\ Lett. \issue(#1,#2,#3)}
\def\MPL(#1,#2,#3){Mod.\ Phys.\ Lett. \issue(#1,#2,#3)}
\def\RMP(#1,#2,#3){Rev.\ Mod.\ Phys. \issue(#1,#2,#3)}
\def\SJNP(#1,#2,#3){Sov.\ J.\ Nucl.\ Phys. \issue(#1,#2,#3)}
\def\CPC(#1,#2,#3){Comp.\ Phys.\ Comm. \issue(#1,#2,#3)}
\def\IJMPA(#1,#2,#3){Int.\ J.\ Mod. \ Phys.\ A \issue(#1,#2,#3)}
\def\MPLA(#1,#2,#3){Mod.\ Phys.\ Lett.\ A \issue(#1,#2,#3)}
\def\PTP(#1,#2,#3){Prog.\ Theor.\ Phys. \issue(#1,#2,#3)}
\def\RMP(#1,#2,#3){Rev.\ Mod.\ Phys. \issue(#1,#2,#3)}
\def\NIMA(#1,#2,#3){Nucl.\ Instrum.\ Methods \ A \issue(#1,#2,#3)}
\def\EPJC(#1,#2,#3){Eur.\ Phys.\ J.\ C \issue(#1,#2,#3)}
\def\RPP (#1,#2,#3){Rept.\ Prog.\ Phys. \issue(#1,#2,#3)}
\def\PPNP(#1,#2,#3){ Prog.\ Part.\ Nucl.\ Phys. \issue(#1,#2,#3)}
\newc{\PRDR}[3]{{Phys. Rev. D} {\bf #1}, Rapid  Communications, #2 (#3)}

\def\PLB(#1,#2,#3){Phys.\ Lett.\ B  \iss(#1,#2,#3)}
\def\JHEP(#1,#2,#3){JHEP \iss(#1,#2,#3)}

\vspace*{\fill}
\vspace{-1.5in}
\begin{flushright}
{\tt HRI-RECAPP-2014-025}
\end{flushright}

\begin{center}
{\Large \bf Gluino mass limits with sbottom NLSP in coannihilation scenarios }
  \vglue 0.4cm
Arindam Chatterjee $^{a}$\footnote{arindam@hri.res.in}
Arghya Choudhury $^{a}$\footnote{arghyachoudhury@hri.res.in},
Amitava Datta$^{b}$\footnote{adatta@iiserkol.ac.in},
Biswarup Mukhopadhyaya $^{a}$\footnote{biswarup@hri.res.in}

\vskip 0.3cm
{$^a$
 Regional Centre for Accelerator-based Particle Physics, \\
Harish-Chandra Research Institute, Allahabad 211019, India
}
\\
{$^b$
Department of Physics, University of Calcutta, 92 A.P.C. Road, 
Kolkata 700 009, India
}\\

	  \end{center}
	  \vspace{.1cm}

\vspace{+1cm}

\begin{abstract}

In view of the recent interest in the pMSSM with light third generation 
squarks, we consider a hitherto neglected scenario where the lighter 
bottom squark ($ \wt{b}_1$) is the next lightest supersymmetric particle (NLSP) 
which co-annihilates with the lightest supersymmetric particle (LSP), 
the dark matter (DM) candidate.
Since the co-annihilation cross section receives 
contributions from both electroweak and strong vertices, it is relatively 
large. As a result relatively large NLSP-LSP mass difference (25 - 35 GeV) 
is consistent with the PLANCK data. This facilitates the LHC signatures of 
this scenario. We consider several variants of the sbottom NLSP scenario 
with and without light stops and delineate the parameter space allowed by 
the PLANCK data. We point out several novel signal 
(e.g., $ \wt{t}_{1} \ra \wt{b}_1 W$) which are not viable in the stop 
NLSP scenario of DM production. Finally, we consider gluino ($\wt g$) 
decays in this scenario and using the current ATLAS data in the jets 
(with or without b-tagging) + $\met$ channel, obtain new limits in the 
$m_{\wt{b}_1} - m_{\wt g}$ mass plane. We find that for  $m_{\wt{b}_1}$ 
upto 500 GeV, $m_{\wt g} \geq$ 1.1 - 1.2 TeV in this scenario.

\end{abstract}

\newpage
\setcounter{footnote}{0}
\section{Introduction}

Supersymmetry (SUSY) \cite{susy} is the most popular and  widely studied 
extension of the standard model (SM) of particle physics. 
In the R-parity conserving minimal supersymmetric standard model (MSSM), the lightest 
neutralino is the lightest SUSY particle (LSP) and a viable dark matter (DM) candidate \cite{dmrev1,dmrev2}. 
R-parity makes the LSP stable leading to missing energy ($\met$) signals at the Large Hadron 
Collider (LHC). Right from the beginning of the LHC run both the ATLAS and CMS 
collaborations are looking for SUSY using this feature of the signals in various channels. 
In the absence of any excess events they have put stringent bounds on the masses of 
supersymmetric particles (sparticles) from the 7 and 8 TeV data. 
The limits on the masses of strongly interacting sparticles are the strongest, due to the  
to large production cross-sections \cite{atlas0l,atlas1l,atlas2l,atlas3b,cms_all}. 
For example, at the end of the 8 TeV run, $m_{squark}~(m_{\tilde{q}})=m_{gluino}~ (m_{\tilde{g}})$ 
scenarios are ruled out upto 1.7 TeV \cite{atlas0l}.

 It should be borne in mind that comprehensive SUSY search strategies
 in the upcoming experiments at LHC 13 TeV runs will be designed on
 the basis of the exclusions obtained by the experiments during the
 first phase. It is , therefore, worthwhile to revisit the limits
 critically and find out the models in which some loopholes in
 the current search techniques significantly relax the limits. A case
 in point is the compressed SUSY scenario which can considerably relax
 the limits \cite{compressed}.

In view of the stringent bounds on the first two generation squark
masses, models with light third generation have received much attention
in recent times \cite{stop_pheno, stop_coan, sbottom_coan, sbottom_pheno}. 
It may be recalled that such heavy squarks
offer a way of ameliorating the SUSY flavour \cite{susy_flavour} and 
CP problems \cite{susy_cp}. 
Such scenarios also help in restoring the naturalness of a SUSY model.

In this analysis, we focus on scenarios with a light $\sbot$, both
with and without a $\lstop$, along with a heavier gluino which
exclusively decays into these squarks. We further assume that the remaining 
members of the MSSM spectrum are heavy, except the lightest neutralino which 
lies below the $\sbot$. The $\sbot$ is the next-to-lightest SUSY particle 
(NLSP) in this scenario. 

An added attraction, and at the same time a crucial viability check 
of this spectrum is that LSP - NLSP 
coannihilation can produce the observed DM relic density of the
universe provided their mass difference is small.  A large number of
phenomenological analyses have already addressed various  topics/issues
related to stop searches in different channels \cite{stop_pheno, stop_coan} and
investigated DM production via $\lstop$ - $\lspone$ coannihilation
\cite{stop_coan} extensively in the context of LHC. On the other hand
the sbottom NLSP scenario, which also has the potential of explaining
the observed relic density, has not yet received due attention and has
been addressed by a relatively small number of analyses only 
\cite{sbottom_coan,sbottom_pheno}. 
Moreover, most of the analyses predating the first phase of
the LHC experiments were restricted to mSUGRA motivated models with
non-universal boundary conditions. Since the SUSY breaking mechanism is
essentially unknown, our emphasis will be on the $\sbot - \lspone$
coannihilation scenario in the phenomenological minimal supersymmetric
standard model (pMSSM) \cite{pmssm} constrained by the LHC data.  Thus, clear
perspectives are expected to emerge from this analysis on (a) the viability 
of  sbottom-neutralino coannihilation (with or without the assistance of stops)
from the standpoint of relic density, (b) the spectacular signatures, not 
viable in the $\lstop$-NLSP scenario, expected in the phase 2 of the LHC 
experiments and c) the new constraints in the $\sbot - \gl$ sector in the 
$\sbot - \lspone$ co-annihilation scenarios using the current LHC data.

It should be noted in addition that one of the scenarios investigated
here, namely the one with just a (right) $\sbot$ and both stop mass
eigenstates heavy, has an appeal from the viewpoint of the lighter
neutral Higgs mass.  It is known that pushing this mass upto $\approx$
125 GeV becomes less troublesome if the stop mass(es) and the
trilinear SUSY breaking term is large.  It therefore helps to have a
situation where both of the stop eigenstates participate in raising
the Higgs mass, while a light $\sbot$ lies close to the lightest
neutralino and and facilitates co-annihilation.

Before delving into our main analysis it is worthwhile to review the
existing LHC limits on $m_{\lstop}$ and $\msbot$ from direct search of
these sparticles.  If $\Delta m_{\lstop} = \mlstop - \mlspone$ is
relatively large so that the decay $\lstop \ra t \lspone$ decay is allowed, 
stop masses in the range 210 - 650 GeV are excluded 
for $\mlspone \lsim$ 30 GeV \cite{stop_atlas}. As $\mlspone$ increases 
the limits become weaker. 
Similarly from sbottom pair production $\msbot < $ 620 GeV is
excluded at 95\% CL for $\mlspone \lsim $150 GeV \cite{sbottom_2b}. 
However, for $\mlspone \gsim$  250 GeV there is no limit on $\mlstop$ 
and $\msbot$. 

None of the above  scenarios, however, is consistent with the observed DM 
relic density of the universe. To explain the correct relic density, we need small 
mass difference between LSP and the NLSP (stop/sbottom) and a different technique for
the NLSP search is called for.
The limits on masses on third generation squarks are remarkably weaker in such
coannihilating scenarios.  From direct stop pair production the latest
bound on $\mlstop$ is 240 GeV \cite{sbottom_mono} for $\mlstop -
\mlspone <$ 80 GeV.  If the stop and the lightest neutralino are
almost degenerate then stop masses up to 260 GeV are excluded from
ATLAS search with a `monojet like' topology \footnote{Here one
depends on a hard ISR jet in enhancing the signal. However, the 
signal with low jet multiplicity may contain more than one jet. }
\cite{sbottom_mono}. Very recently using this search channel and 20
$\ifb$ data at 8 TeV, $\msbot$ below 250 GeV is excluded for small
$\Delta m_{\sbot} = \msbot - \mlspone$ \cite{sbottom_mono}. 

As the gluino pair production cross-sections is the largest, gluino
decays into the third generation squarks are likely to probe larger
ranges of $\mlstop$ and/or $\msbot$ in the coannihilation scenario.
Many groups have looked for gluino decay signatures in presence of light third
generation squarks \cite{gluino_stop_sbot}. In such cases gluino
decays into $\lstop t$ and/or $\sbot b$. For coannihilating scenarios
$\lstop$ decays into $c \lspone$ and $\sbot$ decays into $b \lspone$
with 100\% BRs.  However the $c$ or $b$ jets coming from $\lstop$ or
$\sbot$ will be softer and the limits on the gluino masses could be
degraded. For example, if all three generations squarks are much
heavier than the gluino, then BR($\gl \ra q \bar q \lspone$) is 100\%
where $q$ is any  quark. In this case gluino masses below 
1.4 TeV is excluded for $\mlspone \lsim$ 200 GeV from $jets +\met$ 
channel \cite{atlas0l}. On the other hand assuming $\mlstop - 
\mlspone$ = 20 GeV which is relevant for the $\lstop$ - LSP 
coannihilation and Br($\gl \ra \lstop t \ra c t \lspone$) = 100\%, the 
limit reduces to 1150 GeV \cite{atlas0l}. The ATLAS collaboration have 
also updated their search for NLSP sbottom scenario in the $0-1l + 3b-jets
+ \met$ channel \cite{atlas3b} with Br($\gl \ra \sbot b \ra b b
\lspone$) = 100\% and have excluded $\mgl$ below 1.2
TeV upto $\msbot$ = 1.0 TeV.. But the model is in conflict with the DM relic
density as the LSP mass is fixed at  60 GeV. 

In this paper we have revisited the last analysis focusing on $\Delta
m_{\sbot}$ appropriate for the coannihilation scenario, and
checked whether this relaxes the gluino mass limit. We have considered
several variants of the sbottom NLSP scenario characterized by the
presence or absence of a $\lstop$. In contrast, 
the experimental searches have considered gluino decays either into the
$\lstop$ or the $\sbot$ channel. But if both $\lstop$ and $\sbot$ are
light, then, depending on their mass difference, the gluino may decay into
both the channels with sizable BRs. If, for example, the stop exclusively
decays into a soft charm jet the gluino mass limit may suffer further
suppression. Moreover the presence of an additional light $\lstop$ predicts novel signal for the LHC phase 2 experiments as we shall elaborate below.

An optimized search strategy for the $\sbot$ coannihilation scenario
in the LHC phase 2 experiments calls for immediate attention. In this
paper we use the existing ATLAS search results to obtain the best
limits available at the moment in this scenario with/without a light
stop.  We also compare these results with the ones obtained by a
generic jets + $\met$ search for strongly interacting sparticles.


In Sec. 2, we discuss the variants of the scenario studied here, and identify 
the regions of the parameter space which are consistent with the observed relic 
density as well as direct DM search experiments. We also identify the spectacular signatures in the $\sbot$-NLSP scenario in presence of a relatively light $\lstop$. The new constraints in the $m_{\sbot}$ - $\mgl$ plane in different scenarios using the LHC Run I data
are computed in Sec. 3. We summarize our main conclusions in Sec. 4.


\section{Co-annihilating $\tilde{b}_1$ and $\lspone$ relic abundance: with and
 without light stop:}

We explore a (simplified) pMSSM scenario  with a bino--like 
LSP ($\lspone$) and $\sbot$ as NLSP. Since the bino couples to the other 
(s)particles only through $U(1)_y$ gauge interaction, the annihilation cross-section 
into standard model particles is known to be small. As a consequence, a 
(pure) bino--like $\lspone$ decouples from the thermal soup sufficiently early, 
and therefore has a larger thermal relic abundance compared to the present measured 
relic abundance $\Omega h^2 =0.112 $ \cite{planck} over large region of the 
parameter space. 

However, when there is a scope for co-annihilation with the NLSP the DM and the co-annihilating 
sparticle ($\sbot$ in this case) remain in relative thermal equilibrium for a longer 
period of time through $\lspone ~ SM \leftrightarrow \sbot~ SM'$, where $SM$ and $SM'$ denote 
two Standard Model particles which are assumed to be in thermal equilibrium 
and therefore abundant. 
 In the context of a bino--like $\lspone$, the $s$-channel $\sbot$ 
 exchange process $\lspone~b \leftrightarrow \sbot g$, and the $t/u$-channel $b$ exchange 
processes $\lspone~ g \leftrightarrow b \sbot$ are examples of this. 
Thus, larger (co-)annihilation cross-section of $\sbot$ also implies a lower thermal relic 
density of the $\lspone$. Of course, these interactions eventually decouple and the remaining 
sbottoms decay to $\lspone$. The effect can be captured by \cite{coann} using an effective 
(co-)annihilation cross-section ($\sigma_{eff}$),  
\begin{equation}
\sigma_{eff} = \Sigma_{i,j} \frac{g_i g_j}{g^2_{eff}} 
(1+ \Delta_i)^{3/2} (1+ \Delta_j)^{3/2} e^{-x (\Delta_i+\Delta_j)} 
\sigma_{ij},
\label{eq:coann}
\end{equation}
instead of the annihilation cross-section $\sigma_{ann}$ in the relevant expressions. In the above 
equation, $\{i,j\}$ runs over the list of co-annihilating sparticles, $g_i$ denotes 
the number of degrees of freedom of the $i$-th sparticle, 
$\Delta_i = \dfrac{m_{i}}{m_{\lspone}}-1$, $x = \dfrac{m_{\lspone}}{T} $ and $\sigma_{ij}$ denotes 
the co-annihilation cross-section of $i$ and $j$-th sparticles into SM particles. 
Also,  
\begin{equation}
g_{eff} = \Sigma_i g_i (1+ \Delta_i)^{3/2} e^{-x \Delta_i}.
\end{equation}
Thus co-annihilations are very relevant for sufficiently small $\Delta_i$, i.e. for small mass 
difference between the LSP and the NLSP. 
In this paper we choose the mass difference between $\lspone$ and 
$\tilde{b}_1$ such that sufficient co-annihilation takes place \cite{coann} 
ensuring the correct thermal relic abundance of $\lspone$. 
 Note that, LSP co-annihilation with strongly interacting sparticles 
the involves one strong vertex  leading to large cross-section. 
Consequently to achieve the required $\sigma_{eff}$, the mass difference 
of these  strongly interacting sparticles with the LSP need to be larger 
compared to their electroweak counterparts in a co-annihilation scenario. 
As we will see, while the required mass difference is about 5 - 15 GeV 
for electroweak sparticles (see, e.g., \cite{ew_coan1,ew_coan2}), for (third generation) squarks 
it is about 25 - 35 GeV. This has nontrivial consequences for the observability of $\sbot$ 
coannihilation scenario at the LHC (to be discussed in the next section).

In the pMSSM, the soft-supersymmetry-breaking mass terms for the third generation squarks, 
namely $m_{\tilde{b}_R}$ ($m_{\tilde{t}_R}$) for the $SU(2)_L$ singlets $\tilde{b}_R$ 
($\tilde{t}_R$) and $m_{\tilde{Q_3}}$ for the $SU(2)_L$ doublets are 
free parameters. A large hierarchy among the parameters $m_{\tilde{b}_R}$ ($\tilde{t}_R$) 
and $m_{\tilde{Q_3}}$ ensure that L-R mixing in the sbottom ( stop ) sector is small.
We shall restrict ourselves to such  scenarios only.
We further assume that the first two generations 
of squarks are much heavier and decoupled. The electroweak sparticles other than the 
LSP are assumed to be heavier than $\sbot$ or $\lstop$.
We consider three $\sbot$-LSP coannihilation scenarios characterized by 
different relative magnitudes of the above soft terms leading to potentially  
distinctive LHC signatures. Each scenario is illustrated by a BM point in  Table \ref{tab:bm}
and the corresponding BRs relevant for LHC signatures are tabulated in  Table \ref{tab:br}.
\begin{itemize}
\item  {\bf Gluino-Sbottom-R-Model}: Here the $\tilde{b}_1$ NLSP is 
a $SU(2)_L$ singlet with a small  $m_{\tilde{b}_R}$. The parameters
$m_{\tilde{Q_3}}$ and $m_{\tilde{t}_R}$ are assumed to be too large to 
affect the gluino decays.  
Benchmark point (BMP) BMP-A in Table \ref{tab:bm} provides an example of this scenario.


\item  {\bf Gluino-Stop-Sbottom-R-Model}: In this model the NLSP is dominantly ${\tilde{b}_R}$ as in 
the previous case. However, there is also a R-type light $\lstop$.  In contrast to the 
previous scenario, $\tilde{g}$ decays to both $b\sbot$ and $t \lstop$ channels 
with sizable BRs.  
Depending on the mass splitting between $\sbot$ and $\lstop$, there are three 
possible decay modes of $\lstop$: $c \lspone$, $t \lspone$ and $\sbot W$. They
are tabulated as BMP-B1, BMP-B2 and BMP-B3 respectively in
Table \ref{tab:bm}. 

If the mass difference between $\sbot$ and $\lstop$ is assumed to be significantly 
smaller than $m_W$, $\lstop$ decays dominantly into the loop induced mode 
$c \lspone$. As a result the number of taggable hard jets in the signal decreases 
and this scenario is expected to yield the weakest limits on $\mgl$. Moreover, 
both $\lstop$ and $\sbot$ can contribute to DM production if their mass difference 
is 10 GeV or so. BMP-B1 in the table illustrates this case. 
However, if $\mlstop \gsim m_t+ \mlspone$, then $\lstop \ra t \lspone$  dominates; BMP-B2 
illustrates this scenario. Finally, for $m_W+ \msbot \lsim \mlstop \lsim m_t+ \mlspone$, 
$\lstop \ra \sbot W$  dominates over the flavor-violating decay mode $c \lspone$ 
provided there is a tiny L-R mixing in the stop and sbottom sectors. This is depicted by  BMP-B3. 
Note that in this case  the L-components in  $\lstop$ and $\sbot$ are respectively enhanced  
to about $5\%$ and $0.2\%$ only by adjusting the parameters in the stop and sbottom mass 
matrices. However, if both the mode $\lstop \ra t \lspone$ and  $\lstop \ra \sbot W$ are kinematically allowed the 
mixing angle suppressed latter mode is not competitive. In the presence of large 
mixing in the sbottom and/or stop sector the above classifications become somewhat 
blurred. In this case, e.g, both  $\lstop \ra \sbot W$ and $\lstop \ra t \lspone$ 
may be observed, if kinematically allowed. In fact the relative rates of these two events may 
provide some information on L-R mixing in the squark sector.


 \item  {\bf Gluino-Stop-Sbottom-L-Model}: In this case only $m_{\tilde{Q_3}}$ is relatively light. The $\sbot$ NLSP is dominantly 
an  $SU(2)_L$ doublet i.e, of L-type. In the 
limit of small $L-R$ mixing, $\tilde{t}_1$ is heavier than $\tilde{b}_1$ due to 
larger contributions from the D-terms and F-terms. The mass difference between
$\lstop$ and $\sbot$ is fixed and only allows the  decay $\tilde{t}_1 \rightarrow \tilde{b}_1 W^+$ which occurs with 100 \%. The BMP-C in Table \ref{tab:bm} illustrates this case. 
\end{itemize}

\begin{table}[ht!]
\begin{center}
\begin{tabular}{|c|c|c|c|c|c| } \hline
Parameter	& BMP-A	&BMP-B1	&BMP-B2	&BMP-B3	& BMP-C  \\
\hline
\hline
$\mlspone$    & 300    & 306	&305	& 305	&305    \\    
$\mgl$        & 1236   & 1259  &1273	& 1270	& 1310   \\
$\msbot$      & 325    & 335   &333	&333 	& 334    \\
$\mlstop$     & 1558   & 345   &507	& 468	& 455    \\
\hline
\hline
$\Omega_h^2$  & 0.11   & 0.11  & 0.11	& 0.11	& 0.12   \\
\cline{1-6}
\end{tabular}
\end{center}
\caption{ {\it \small Mass spectra for different benchmark points. 
BMP-A,B(1,2,3),C represent three different scenarios Gluino-Sbottom-R-Model, 
Gluino-Stop-Sbottom-R-Model and Gluino-Stop-Sbottom-L-Model 
respectively (see text for details). All the masses are in GeV.} }
\label{tab:bm}
\end{table}

The BRs relevant for LHC signatures are tabulated in Table \ref{tab:br}.
In all scenarios, Br($\sbot \ra b \lspone$) is 100\%, as $\sbot$ is the NLSP. 
In the scenarios with a light $\lstop$ the $\tilde{g}$ decays to both $b\sbot$ 
and $t \lstop$. For the scenarios represented by BMP-B1, BMP-B3 and BMP-C the 
BRs of the two modes are approximately the same irrespective of $\mlstop$. In BMP-B2 
the BRs of the final states with $\lstop$ decreases as $\mlstop$ approaches $\mgl$  
and the scenario is indistinguishable to Gluino-sbottom-R model.  We have chosen 
BMP-B2 such that $\mlstop$ the $\gl$ decays to both modes with approximately 
the same BR. As discussed above, in the limit of small L-R mixing  $\lstop$ decays 
into a single channel channel with almost 100 \% BR for $\mlstop$ in a specific range. 
BMP-B1 and BMP-B2 represent scenarios with qualitatively different stop decays and can 
in principle be distinguished from other scenarios. On the other hand BMP-B3 and 
BMP-C are indistinguishable so far as stop decays are concerned.

\begin{table}[ht!]
\begin{center}
\begin{tabular}{|c|c||c|c|c||c|} \hline
{Decay Modes} 			& BMP-A &BMP-B1	&BMP-B2	&BMP-B3	& BMP-C  \\
\hline
\hline
$\gl		 \ra b \sbot 	$& 100	&52.3	&52.2	&50.8	& 54.4	\\ 
$\quad	 	 \ra t \lstop  	$& -	&47.7	&47.8 	&49.2	&45.6\\
\hline   
\hline 
$\sbot		 \ra b \lspone 	$& 100	&100	&100	&100	& 100   \\
\hline
\hline
$\lstop		\ra c \lspone 	$& -   &97    	&-	&-	&-   \\
$\quad 		\ra W \sbot	$& -   &$3^{*}$&1.4	&100	&100	\\
$\quad 		\ra t \lspone	$& -   &-	&98.6	&-	&-	\\
\hline
\end{tabular}
\end{center}
\caption{ {\it  \small Branching Ratios - BRs ($\%$) of decay modes of $\gl$, 
$\sbot$ and $\lstop$ for benchmark points. $*$ For BMP-B1  
small BRs of $\lstop \ra W \sbot$ is coming from virtual $W$.} }
\label{tab:br}
\end{table}

As already noted no dedicated search for the $\sbot$ NLSP has so far been carried 
 out at the LHC. In Sec. 3 we shall focus on new  constraints on $\msbot$ and 
$\mgl$ using the available data from phase  1 of the LHC run. However it should 
be emphasized that spectacular signatures of gluino decays in the $\sbot$ NLSP 
scenario with a light $\lstop$ will be worth searching at the LHC during phase 2. 
Final state topologies like $ 2b W  \met$, $b t W \met$ 
or $b t  \met$ are not viable in the $\lstop$ - NLSP scenario.   

We ensure 
that the spectrum, generated by \texttt{SuSpect} \cite{suspect}, 
contains a CP-even (light) Higgs with mass $125 \pm 3$ GeV, as required by 
the present LHC data \cite{higgs1,higgs2}. 
The relevant modes for $\gl$, $\sbot$ and $\lstop$ decays with their branching ratios (BRs) 
for the benchmark points are shown in Table \ref{tab:br}. 
We have used \texttt{SUSYHIT} \cite{susyhit} to obtain the decay widths and branching 
ratios in various channels. We have computed the DM relic density using \texttt{micrOMEGAs-3.6.7}.
\cite{micromega3}.

For the \textit{Gluino-Sbottom-R-Model}, in estimating the thermal 
abundance, three processes play major roles. The first one is $\sbot \sbot^* 
\rightarrow g g$, which contributes dominantly to $\sigma_{eff}$. This process receives 
contributions from four-point (gauge) interaction, $s$-channel $g$ exchange, and 
$t/u$-channel $\tilde{b}$ exchange processes where the first two channels contribute most. 
In addition, there can be annihilation via $\lspone \sbot \rightarrow gb$ and $\sbot \sbot \rightarrow bb$. 
Note that $\lspone$ annihilation is a pure EW process and hence its contribution 
remains small. While $s$-channel $b$ exchange and $t/u$-channel $\sbot$ exchange 
contributes to the former one; the latter one is mediated by  $\gl$ and neutralinos 
(mostly the bino--like one) in $t/u$-channel. For our BMP-A, these three processes 
contribute $63\%,~ 18\%$ and $11\%$ respectively. Although, because of large $\mgl$ 
mass, the contribution from the third process is rather small for smaller $\lspone$ 
masses, it increases for larger $\lspone$ masses contributing about $20\%$ for 
$m_{\lspone} \sim 600 $ GeV. 

In the \textit{Gluino-Stop-Sbottom-R-Model}, 
for BMP-B1, $\sbot \sbot^* \rightarrow g g$ contributes about 33\%; $\lstop \sbot \rightarrow gb$ 
and $\lspone \tilde{t}_1 \rightarrow gt$ contribute about 15\% each; and $\lspone \lspone 
\rightarrow t \bar{t}$ contributes about 11\%. The latter receives contributions from 
$\tilde{t}_1$ exchange $t/u$-channel processes, and is unsuppressed due to the large 
top mass which helps in evading large chirality suppression. Any other channel contributes 
less than 10\% for this chosen benchmark point. In BMP-B2 and BMP-B3, $\lstop$ is quite heavy compared to $\lspone$. 
Consequently, the leading co-annihilation channels involve $\sbot$ only, and their 
relative contributions are similar to that in Gluino-Sbottom-R-Model (BMP-A). However, 
since $\lstop$ is quite light compared to that in BMP-A, $\lspone$ annihilation into 
$t \bar{t}$ (mediated via $\lstop$ in $t/u$-channel) contributes about $5-6\%$ in 
both these cases. 

Finally, in the \textit{ Gluino-Stop-Sbottom-L-Model}, the leading contribution 
comes from $\sbot \sbot^* \rightarrow g g$, which gives about 34\% of the total 
coannihilation rate for BMP-C. Since $\sbot$ is L-type, large D-term contribution  
together with contributions from the F-term, leads to large $\sbot \sbot^* \rightarrow hh$  
cross-section. Along with four-point vertices, $s$-channel $h$ mediated process and 
$t/u$-channel $\sbot$ mediated processes contribute to this channel. 
Altogether, its contribution is about 22\%. This is followed by the cross-sections for 
$\sbot \sbot \rightarrow bb$, and other channels which contribute less than 10\% 
for BMP-C. Since $m_{\tilde{t}_1}$ is quite large (due to large D-term and F-term 
contribution, in the no-mixing limit), co-annihilation with $\tilde{t}_1$ does not take 
place.

\begin{figure}[!htb]
\begin{center}
{\includegraphics[angle =270, width=0.8\textwidth]{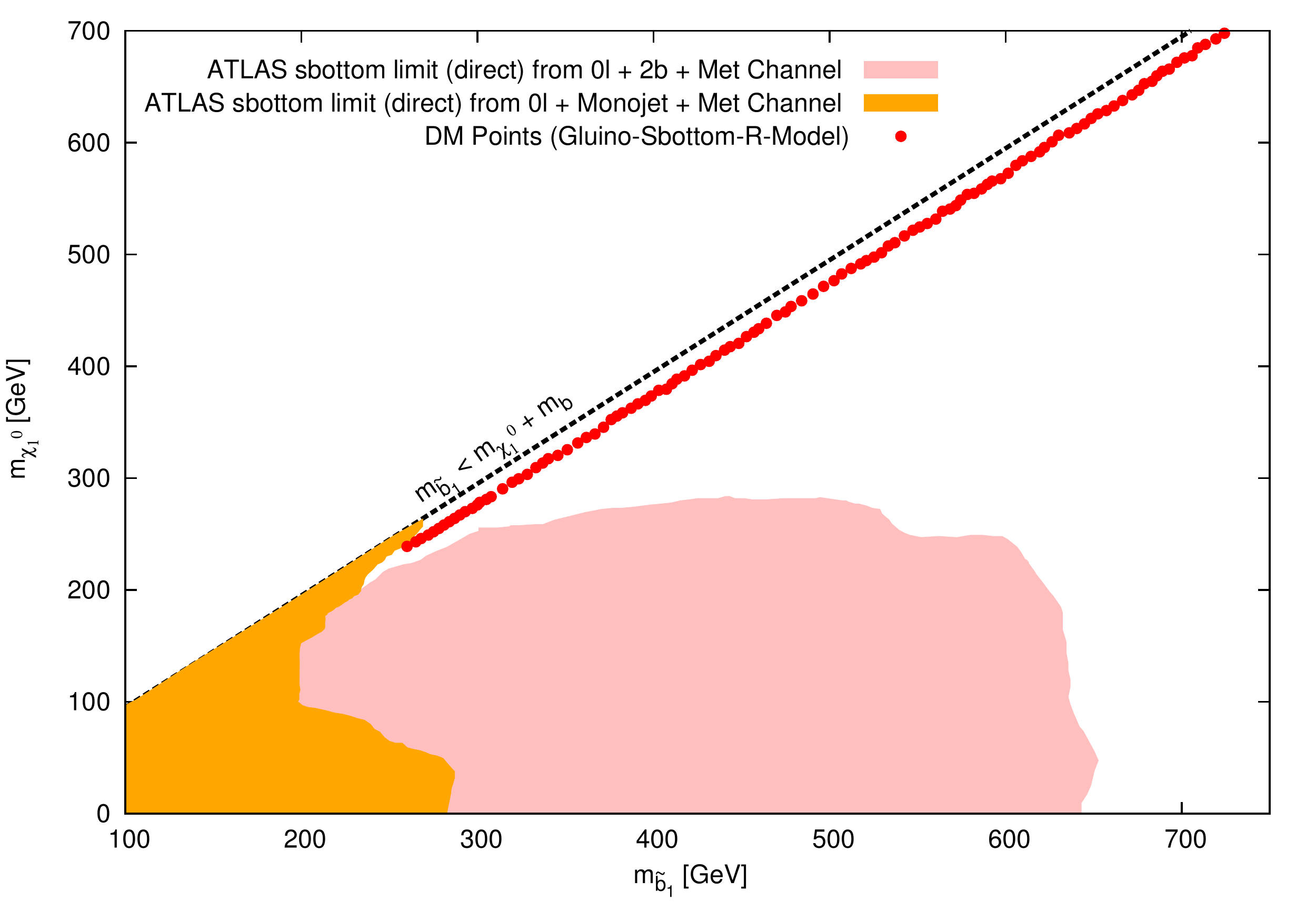} }
\caption{ {\it \small  The red points give correct thermal relic density of $\lspone$ allowed 
by PLANCK \cite{planck} in Gluino-Sbottom-R-Model. Limits on $\msbot$ from direct 
$\sbot$ pair production from $0l$ + monojet + $\met$ search \cite{sbottom_mono} and 
$0l$ + 2b + $\met$ search \cite{sbottom_2b} by ATLAS are presented by orange 
and pink shaded region respectively. }}
\label{fig1}
\end{center}
\end{figure}

In Fig.\ref{fig1}, the red points show the region where $\lspone$ has the right thermal 
relic abundance in \textit{Gluino-Sbottom-R-Model}. The mass difference with $\tilde{b}_1$ does 
not change significantly over a large region of $m_{\lspone}$; it varies between about 25 - 35 GeV. 
Since strong interaction processes involving $\sbot$ always contribute dominantly to the relic 
density, such a minor variation of $(m_{\sbot}-m_{\lspone})$ is well anticipated, 
especially since $\sigma_{eff}$ depends exponentially on the same. 
The figure further shows the present ATLAS limit on $m_{\tilde{b}_1}$ derived from 
$0 l + 2 b+  \slashed{E}_T$ and $0l+1j+ \slashed{E}_T$ (compressed scenario) for direct $\sbot$ 
production \cite{sbottom_mono,sbottom_2b}.

In \textit{Gluino-Stop-Sbottom-R-Model} $\tilde{t}_1$ ($R$-type) also contributes to the 
co-annihilation processes; for BMP-B1 its contribution is about 30\%.  Correspondingly, 
the mass difference $(m_{\sbot}-m_{\lspone})$ is raised by about 4 GeV compared to that in 
\textit{Gluino-Sbottom-R-Model} scenario. The mass difference varies between about 28-35 GeV
over a large range (up to about 800 GeV) of $m_{\lspone}$ in this scenario too.

Green (blue) Points in Fig. 2a(2b) represent the parameter space with correct thermal relic 
density allowed by PLANCK in the Gluino-Stop-Sbottom-R(L)-Model. Orange and pink shaded 
regions are excluded from  direct $\sbot$ pair production data in 
$0l$ + 2b + $\met$ channel \cite{sbottom_2b} and $0l$ + monojet + $\met$ channel 
\cite{sbottom_mono} by ATLAS.

Note that, we have used gluino masses around 1200 GeV to estimate 
the relic density. However, the $t$ (and $u$) channel gluino exchange 
processes $\tilde{b}_1 \tilde{b}_1 \rightarrow b ~b$ contributes only  
about $10 -25\%$ to $\sigma_{eff}$ during DM freeze-out for our benchmark points. 
 \footnote{ 
 In $\tilde{b}_1 \tilde{b}_1^* \rightarrow b \bar{b}$ $s$ channel $g$ exchange 
 contributes dominantly along with $t$ (and $u$) channel $\tilde{g}$ exchange 
 process. However, this channel contributes less than 1\% to the thermally 
 averaged annihilation cross-section.} 
Even a large increment in $\mgl$, therefore, can be compensated by a corresponding 
reduction in the mass of the NLSP by a few GeV, retaining the correct thermal 
relic abundance. We have checked this numerically using \texttt{micrOMEGAs} \cite{micromega3}. 
 \footnote{Note that, the mass difference between the LSP and the NLSP affects 
  the contributions from the co-annihilating channels to $\sigma_{eff}$ exponentially.}

\begin{figure}[!htb]
\begin{center}
\subfloat[]{\includegraphics[angle =270, width=0.45\textwidth]{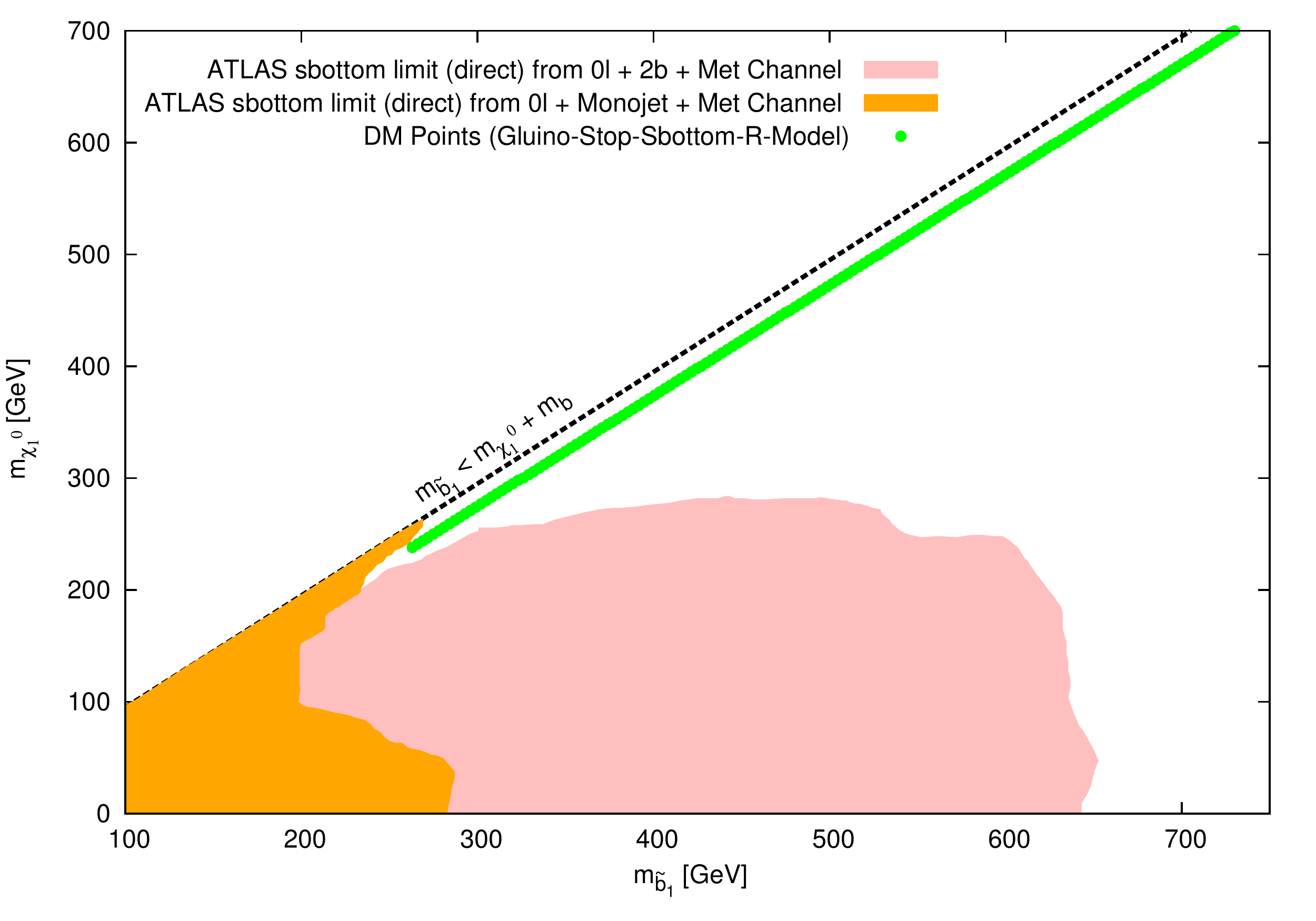} }
\subfloat[]{\includegraphics[angle =270, width=0.45\textwidth]{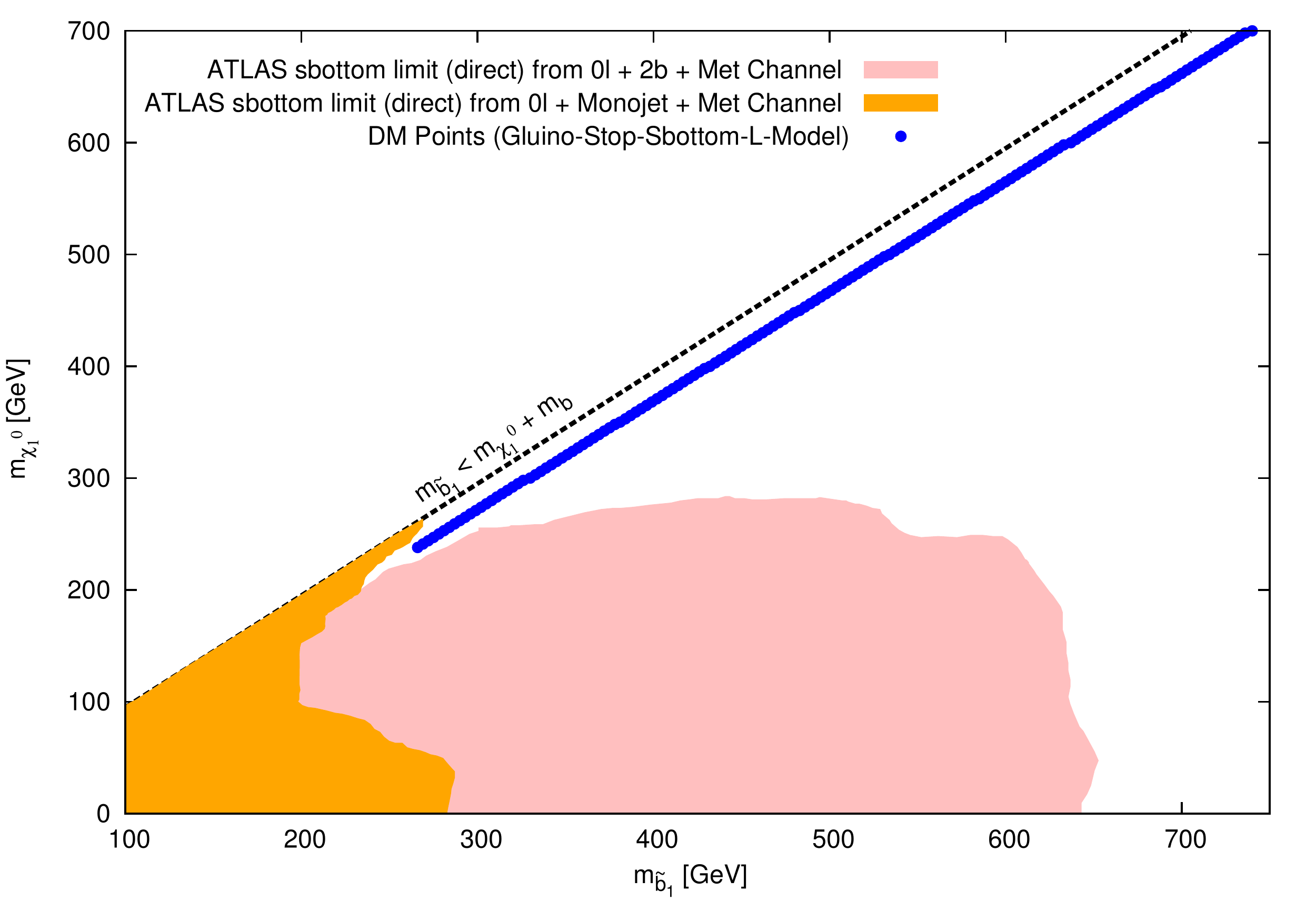} }
\caption{ {\it \small  The green and blue points give correct thermal relic 
density of $\lspone$ allowed by PLANCK \cite{planck} in Gluino-Stop-Sbottom-R-Model (left panel) 
and Gluino-Stop-Sbottom-L-Model (right panel) respectively (see text for details).  
Limits on $\msbot$ from direct 
$\sbot$ pair production from $0l$ + monojet + $\met$ search \cite{sbottom_mono} and 
$0l$ + 2b + $\met$ search \cite{sbottom_2b} by ATLAS are presented by orange 
and pink shaded region respectively. }}
\end{center}
\label{fig1bc}
\end{figure}

It should be further noted that we have assumed $|\mu| =1000 {\rm GeV}$. While reducing 
$|\mu|$ has no effect on the (2-body) decay of $\tilde{g}$; significant bino-higgsino 
mixing in $\lspone$ can be achieved. However, such mixing is quite constrained by 
the present data from \texttt{LUX} \cite{lux}. For a 300 GeV $\lspone$, we found it to 
be at most about 5-10\%. But the details of direct detection cross-section also depend on 
$\tan \beta$ and the heavy Higgs masses. Moreover, uncertainties from the estimation of 
the strange quark content of the nucleon affect the direct detection cross-section 
significantly; assuming the default values of strange quark content at low energies 
may turn out to be an oversimplification \cite{micromega3}. In the present context, its effect on the thermally averaged 
annihilation cross-section can again be compensated by changing the mass of the NLSP by a 
few GeV, thus restoring the correct thermal relic abundance. It may thus be worthwhile to 
consider sbottom NLSP scenarios with low values of $|\mu|$ as well.   

In a nutshell, we have considered various scenarios with a light 
$\tilde{b}_1$ NLSP (and also light $\tilde{t}_1$ in benchmarks BMP-B and BMP-C).  
In all these scenarios, the bino--like $\lspone$ provides a suitable Dark Matter 
candidate. While its dominantly bino--like nature assures that it escapes the present direct searches, 
co-annihilation with $\tilde{b}_1$ plays a crucial role in achieving the right thermal 
relic density. In the next section, we will consider the lower bound on the mass 
of $\tilde{g}$ in these scenarios.


\section{Constraints on the $\sbot-\gl$ sector in the $\sbot$  NLSP scenarios}

In this section we focus on $\gl$ pair production 
in the sbottom NLSP scenarios described in the previous section and obtain the 
exclusion contours in the $\mgl$ - $\msbot$ plane using the LHC data. For each 
point in the plot $\mlspone$ is chosen in the narrow range consistent with the 
observed DM relic density.  We also compute the limits on $\mgl$ for  
different models described in previous section and also for different BMPs 
introduced in Table \ref{tab:bm} and \ref{tab:br}. Since there is no dedicated LHC search for 
this case our aim is to constrain the gluino mass utilizing
the ATLAS model independent cross section bounds (see below) for final states 
similar to but not identical with the ones we are interested in. Our choice is
 guided by the fact that 
the final state coming from $\gl \gl$ pair production in all cases is expected to  
be enriched with $b$-jets and the generic signal will be jets + $\met$.

The ATLAS collaboration has updated their SUSY search results at 8 TeV with 
$\lum \sim$ 20 $\ifb$ data in 
$n$-leptons + $m$-jets (with or without $b$ tagging) + $\met$ channel for different integral values 
of $n$ and $m$ and interpreted the results for various simplified models. 
Here we will concentrate mainly on the  jets (at least $3b$ jets) + 0-1$l$ 
($l$ = $e,\mu$) + $\met$ \cite{atlas3b} signal. The results were interpreted in a 
simplified model with light $\gl$ and $\sbot$  for a fixed LSP mass: $\mlspone$ = 60 GeV.  
We have obtained new constraints for $\Delta m_{\sbot}  \approx$  30 GeV, which the 
hallmark of the $\sbot - \lspone$ coannihilation scenario. We have also considered the 
channel jets (no $b$ tagging) + 0$l$ ($l$ = $e,\mu$) + $\met$  \cite{atlas0l}. 
Although this analysis leads to weaker $\mgl$ limits in most cases of our interest, 
it yields the most stringent bounds for significantly smaller values of $\Delta m_{\sbot}$. 
Such choices, though disfavoured in the $\sbot$ - $\lspone$ co- annihilation scenarios, 
can not be absolutely ruled out in view of the fact that DM production may be due to other mechanisms.

Next we will briefly review the above  analyses by ATLAS. 
SUSY searches in the inclusive jets + $0l$ + $\met$ channel for $\lum$ = 20.3  $\ifb$ at 
8 TeV have been presented in Ref.~\cite{atlas0l}. 
Five inclusive analyses channels, labelled as A to E depending on jet multiplicities (from 
two to six respectively), are introduced. The relevant cuts have been described in 
Table~1 of ref. \cite{atlas0l}.  Each of these channels are further classified as 
`Tight',`Medium' and `Loose' based on the final cuts on the observables $\dfrac{\met}{m_{eff}}$ 
and $m_{eff}$(incl.)
\footnote{$m_{eff}$ is defined as the scalar sum of the transverse momenta of the leading 
N jets which defines the signal region and $\met$. $m_{eff}$(incl.) is defined as the scalar 
sum of the transverse momenta of the jets having $P_T$ greater than 40 GeV and $\met$.}. 
Non-observation of any significant excess in each of these signal region leads to 
an upper limit on the number of events ($N_{BSM}$)from any Beyond Standard Model 
(BSM) scenario. The observed upper limits on $N_{BSM}$ at 95 $\%$ Confidence Level 
(CL) for signal regions SRA-Light, SRA-Medium, SRB-Medium, SRB-Tight, SRC-Medium, SRC-Tight, 
SRD, SRC-Loose, SRE-Loose, SRE-Medium, SRE-Tight are given by 1341, 51.3, 14.9, 
6.7,  81.2,  2.4, 15.5, 92.4, 28.6, 8.3 respectively \cite{atlas0l}. We use these 
(model independent) numbers to obtain new limits on $\mgl$. 


ATLAS collaboration has also reported the search for strong sparticles
in the multi-$b$-jets final states with $\lum$ = 20.1 $\ifb$ at 8 TeV in \cite{atlas3b}. 
Selection criteria for the signal regions are listed in Table 1 and 2 of Ref. \cite{atlas3b}. 
In this analysis both $0l$ (two signal regions) and at least one lepton (one signal region) 
channel are introduced. Signal regions  are characterized by large $\met$ and at least
four ($SR-0l-4j$), six ($SR-1l-6j$) or seven ($SR-0l-7j$) jets which includes at least 
three $b$-tagged jets. Finally, these are further classified as $A/B/C$ depending on 
$\met$ and $m_{eff}$. The absence of any excess led to an upper limits on 
the number of signal events in each of these regions from ATLAS. In particular,  
for signal regions  
SR-0l-4j-A, SR-0l-4j-B, SR-0l-4j-C,  
SR-0l-7j-A, SR-0l-7j-B, SR-0l-7j-C, 
SR-1l-6j-A, SR-1l-6j-B, SR-1l-6j-C 
these upper limits, at 95 $\%$ CL, are given by 4.6, 6.7, 4.8, 15.3, 6.1, 3.9, 6.6, 
3.0, 3.0 respectively.

We adopt different selection criteria for various signal regions discussed above.  
For electron, muon and jet identification, reconstruction, isolation etc, 
we use the ATLAS prescription described in the respective analyses \cite{atlas0l,atlas3b} 
\footnote{Apart from the two analyses reported here we have also implemented 
the constraints from the hard single lepton channel - $1l$ + jets + $\met$ channel \cite{atlas1l} 
and the 2$l$/3$l$ + 0-3 bjets + $\met$ channel\cite{atlas2l} in our code, 
but in most of the cases these channels only impose weaker constraints.}. The $P_T$ 
dependent b-tagging efficiencies provided by ATLAS collaboration in Ref. \cite{btagging} 
have been used in our code. For validation, we reproduced the number of events in 
each of these signal regions, as obtained by ATLAS, for some benchmark 
points in Refs.\cite{atlas0l,atlas1l,atlas2l,atlas3b}.

\begin{figure}[!htb]
\begin{center}
{\includegraphics[angle =270, width=0.8\textwidth]{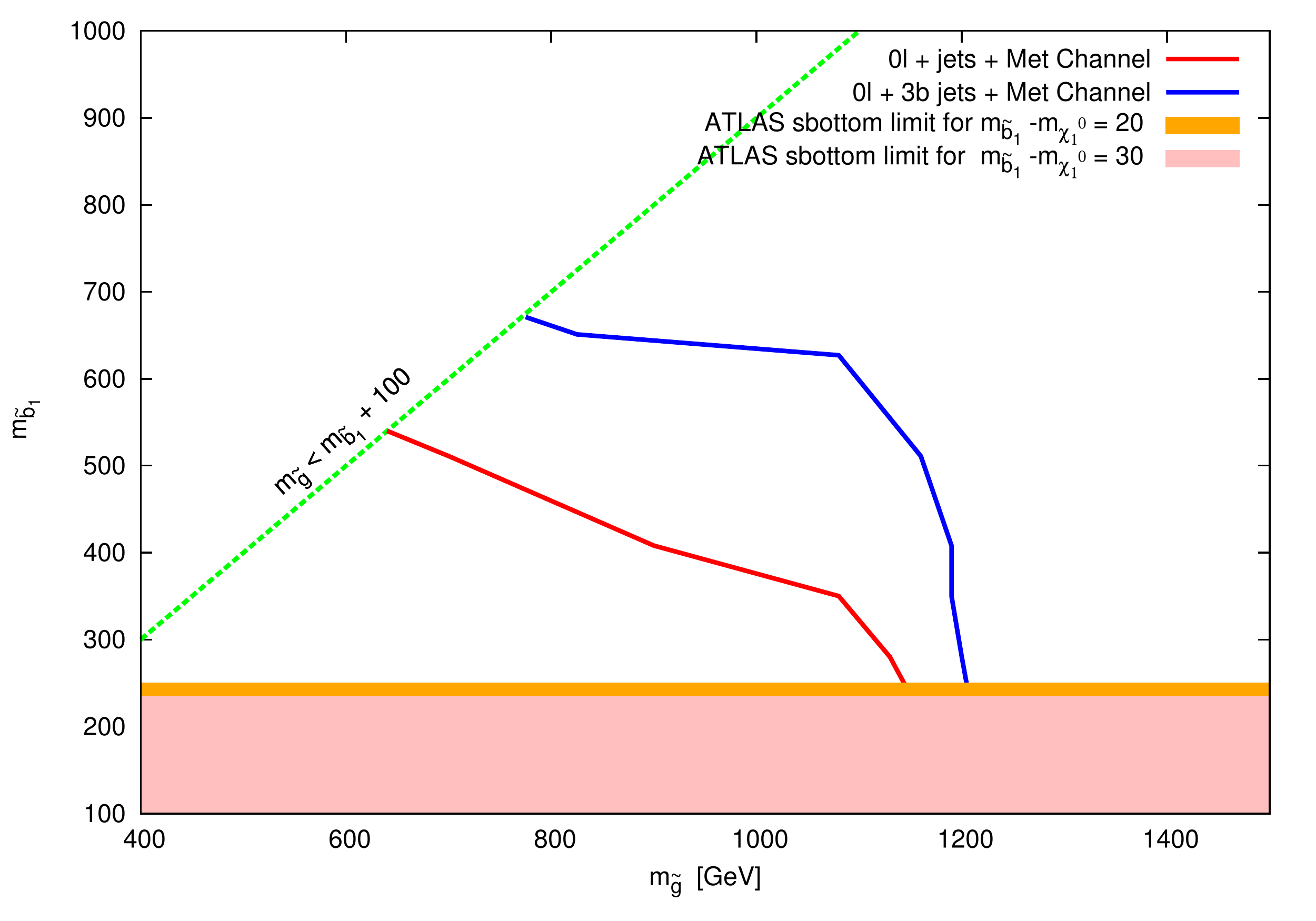} }
\caption{ {\it  \small 
Limits on $\mgl$ in Gluino-Sbottom-R-Model from $\gl\gl$ pair production 
in $\mgl-\msbot$ plane from 
$0l + jets +\met$ \cite{atlas0l} and  $0l + jets (3b) +\met$ channel \cite{atlas3b} 
at 95 \% $CL$ with 8 TeV 20 $\ifb$ ATLAS data. 
The shaded region are excluded from direct $\sbot$ pair production 
\cite{sbottom_mono, sbottom_2b} (see Fig. \ref{fig1})
  }}
\label{fig2}
\end{center}
\end{figure}

We then use PYTHIA (v6.428) \cite{pythia} to generate the signal events in different 
channels from gluino pair production for different scenarios. 
The NLO cross-section for the $\gl \gl$ pair production is computed with  
PROSPINO 2.1 \cite {prospino} using CTEQ6.6M PDF \cite {cteq6.6}.  
Finally we derive the new limits on gluino mass by comparing the computed number 
of events with the corresponding upper limits on $N_{BSM}$ in the relevant signal 
regions. The exclusion regions for Gluino-Sbottom-R-Model, thus obtained, are 
presented in  Fig. \ref{fig2}. 

\begin{figure}[!htb]
\begin{center}
{\includegraphics[angle =270, width=0.8\textwidth]{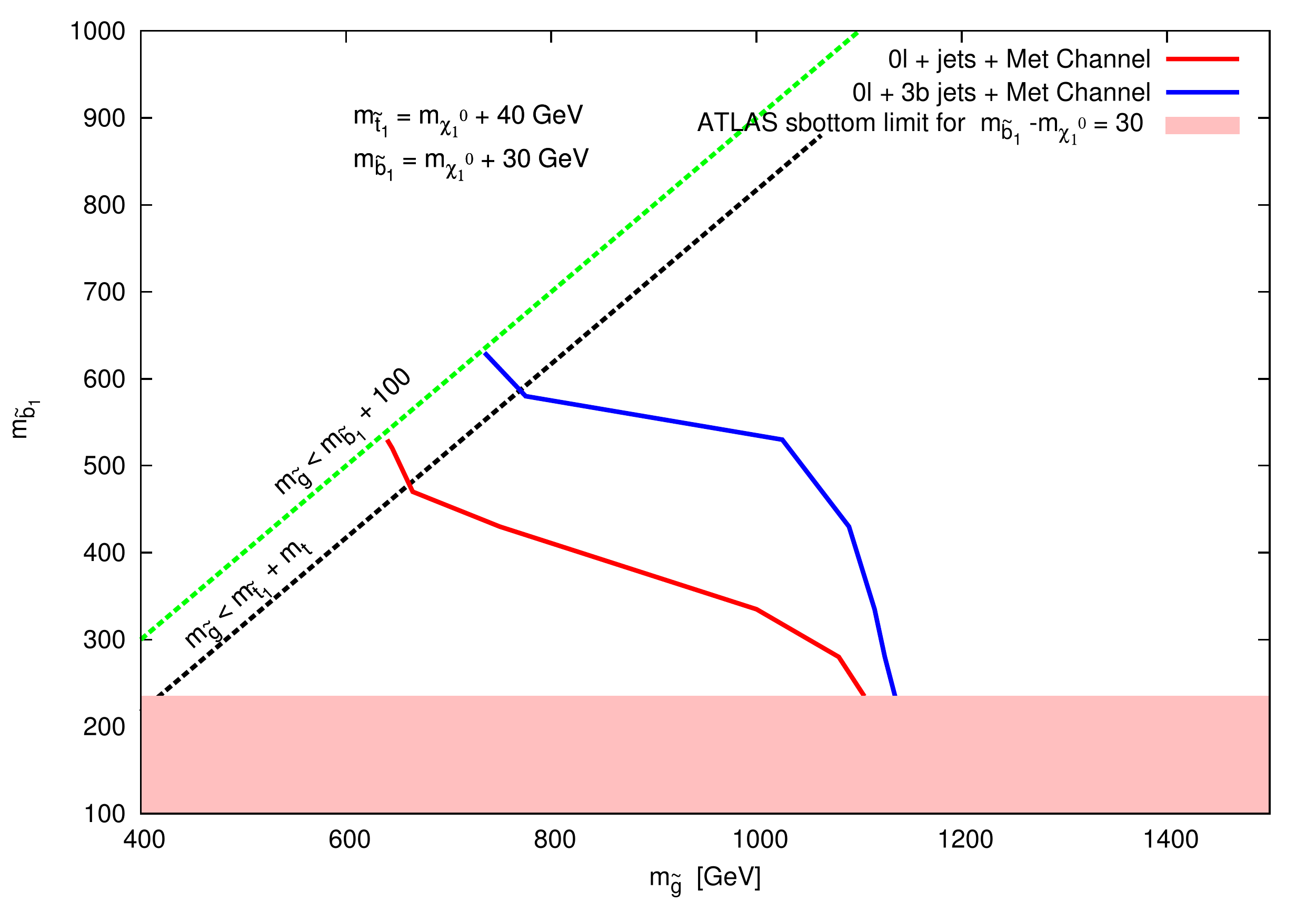} }
\caption{ {\it \small 
Limits on $\mgl$ in Gluino-Stop-Sbottom-R-Model from $\gl\gl$ pair production 
in $\mgl-\msbot$ plane from 
$0l + jets +\met$ \cite{atlas0l} and  $0l + jets (3b) +\met$ channel \cite{atlas3b} 
at 95 \% $CL$ with 8 TeV 20 $\ifb$ ATLAS data. 
Here $BR(\sbot \ra b \lspone)$ = 100\% and $\lstop \ra c \lspone$ dominates 
(for illustration see BMP-B1). 
The shaded region are excluded from direct $\sbot$ pair production 
\cite{sbottom_mono, sbottom_2b} (see Fig. \ref{fig1})
  }}
\label{fig3}
\end{center}
\end{figure}

In Fig. \ref{fig2} each point gives correct DM relic density via sbottom 
coannihilation i.e., mass difference between $\sbot$ and $\lspone$ is in the
range 25 - 35 GeV. We only consider the parameter space $\mgl > \msbot + 100$ GeV. If $\mgl$ is 
smaller then, as discussed in Sec. 2, the gluino contributes to  DM relic production significantly leading
to under abundant relic density. The shaded region in Fig. \ref{fig2} is excluded by direct search for 
sbottom pairs for low $\Delta {\msbot}$ . 

The most stringent bound on $\mgl$ comes from $0l + jets (3b) +\met$ channel \cite{atlas3b} 
-- for $\msbot$ around 500 GeV, gluino mass below 1.1 - 1.2 TeV is excluded.  
The red and blue lines in Fig. \ref{fig2} represents the exclusion contour from 
$0l + jets +\met$ \cite{atlas0l} and  $0l + jets (3b) +\met$ channel \cite{atlas3b}. 
Due to relatively soft b-jets from $\sbot$ decays, 
the signal in the sbottom NLSP scenario is sensitive to signal regions with low jet multiplicity. 
The two most effective signal regions are SRB-Medium \cite{atlas0l} 
and SR-0l-4j-A \cite{atlas3b} which require 3jets and 4jets 
respectively. 
In the Gluino-Stop-Sbottom-R-Model depending upon stop decay modes we consider 
three possibilities (see  and Table \ref{tab:bm} and Table \ref{tab:br}). 
If  the decay $\lstop \ra c \lspone$ is the most dominant mode (see BMP-B1), 
then the gluino mass limits is expected to be rather weak. 
For our computations in this case we assume the 
mass relations $\msbot = \mlspone + 30$ GeV and $\mlstop = \mlspone + 40$ GeV. 
As already noted the relatively light $\lstop$ also contributes  to  DM production.  We present the results in 
Fig.~\ref{fig3} for this model. Gluino limits in this model are weaker than in 
Fig.~\ref{fig2} roughly by 100 GeV in both channels. 
This is due to the fact that the  gluino decays into $\lstop t$  with almost 50\% BRs and effectively 
reduce the $0l$ or $bjet$ signal events. It may be noted that (see  Fig. \ref{fig3}) when 
$\gl \ra \lstop t$ is kinematically disfavoured, then 
the limits tend to increase to some extent.

In the other scenarios where $\lstop \ra t \lspone$ 
or $\lstop \ra W \sbot$ dominates, there are extra sources of jets and tagged bjets. 
As a result the limits on $\mgl$ are pushed up. As the exclusion contour or mass limits lie between the results obtained in 
Fig. \ref{fig2} and Fig. \ref{fig3}, we have not presented any detailed figure for such 
scenarios. Instead we present the limits on $\mgl$ for the 
benchamark points (see Table \ref{tab3}). As expected BMP-A gives the strongest limit $\mgl \gsim$ 1.2 TeV and for 
BMP-B1 it reduced to 1.1 TeV. We note in passing that for BMP-B2, BMP-B3 and BMP-C 
the signal contain multiple $W$s which reduces the limits from 
the $0l +jets +\met$ channel \cite{atlas0l}. In such cases, however, the hard single 
lepton channel ($1l +jets +\met$) \cite{atlas1l} puts  more stringent bounds ($\mgl >$ 1025 - 1050 GeV).
\begin{table}[!htb]
\begin{center}\
\begin{tabular}{||c||c||c||}
\hline
Points		& \multicolumn{2}{c|}{Limit on $\mgl$ (GeV)} 		\\
\cline{2-3}
& $0l +jets+  \met$\cite{atlas0l} 	&$0l +jets(3b) + \met$ \cite{atlas3b}	\\
\hline
BMP-A	 	& 	1070		& 1195	\\
\hline
BMP-B1	 	& 	1000 		&1115	\\
\hline
BMP-B2	 	& 	845 		&1165	\\
\hline
BMP-B3	 	& 	850 		&1135	\\
\hline
BMP-C	 	&	850 		&1150	\\

\hline
\hline
       \end{tabular}\
       \end{center}
           \caption{ {\it \small Limits on $\mgl$ using the ATLAS $0l + jets $ + $\met$ data\cite {atlas0l}, 
 $0l + jets (3b) +\met$ data\cite{atlas3b} for the benchmarks. }}
\label{tab3}
          \end{table}


If $\Delta m_{\sbot}$ happens to be much smaller than the values allowed by the DM relic density constraint, 
then $0l + jets +\met$ channel yields a stronger lower bound on $\mgl$. For example, if $\Delta m_{\sbot}$ = 10~GeV, 
a choice which cannot be strictly ruled out if the possibility of non-supersymmetric DM is taken into account, 
the bound is about 1 TeV in this channel. The corresponding limit from the $0l + jets (3b) +\met$ data is 
much weaker (775 GeV).

In the $\lstop$ NLSP scenarios with correct DM relic density, the decays $\sbot W$ and $t \lspone$ are not allowed. 
However there are sufficiently large parameter space in the $\sbot$ NLSP scenario where these decays are allowed.
These decays may provide  novel signatures of this scenario during the LHC phase 2 experiments.

\section{Conclusion}

In view of the current interest in SUSY with light third generation squarks, 
we have studied the co-annihilating sbottom NLSP scenario and have delineated 
the parameter space consistent with the DM relic density constraint (see Fig. \ref{fig1}) 
when the NLSP is of R-type with both of the stop mass eigenstates being rather 
heavy (the Gluino-Sbottom-R model). It easier to fit the observed Higgs mass 
in this case. It should also be stressed that in this scenario the co-annihilation 
cross section is relatively large since strong interactions partially contribute 
to it. As a result somewhat large (25 - 35 GeV) $\Delta m_{\sbot}$ 
is consistent with the PLANCK data ( see Eq. \ref{eq:coann}). 
This enhances the observability of the LHC signatures.

However, there are other options with an additional light $\lstop$ like the 
Gluino-Sbottom-Stop-R model and the  Gluino-Stop-Sbottom-L-Model (Sec. 2). 
The allowed range of $\Delta m_{\sbot}$ is very similar to the one in the 
previous model (see Fig. 2). Some  of the models predict spectacular 
LHC signatures with multiple tagged b-jets, $W'$s and  reconstructable $t's$ over 
a large parameter space which should be searched for during phase 2 (Sec. 2). 
These signals are not viable in the $\lstop$ - NLSP scenario.
The fraction of events with reconstructed $W$, top etc. can reveal  
the underlying scenarios  including some  hints on the mixing angles in the  stop and sbottom sectors.

We note in passing that the $(g-2)_{\mu}$ anomaly \cite{muong2} can be easily resolved within
the framework of the $\sbot$ NLSP scenario with relatively heavy electroweak sparticles.  
For example, if the sleptons are heavy, $\mlspone \gsim$ 100 GeV, the constraint on the  
lighter chargino ($\chonepm$) mass from LHC is very weak and a large parameter space is  
consistent with the $(g-2)_{\mu}$ constraint. For lighter LSP there are stronger LHC 
constraints on the chargino mass; yet a reasonable parameter space compatible with the 
$(g-2)_{\mu}$ constraint is allowed (for a recent discussion see Fig. 8 and Sec. 3.4 of 
Ref.~\cite{ew_coan2}). In the Gluino-Sbotom-R  scenario the allowed $\sbot$-NLSP masses 
are bounded from below by the LHC lower limit (260 GeV) from direct searches of coannihilating 
sbottom squarks. On the other hand $m_{\sbot}$ is bounded from above by the maximum 
chargino mass determined by the $(g-2)_{\mu}$ constraint, the LSP mass and the LHC searches 
for the electroweak sparticles. The same conclusion holds for Gluino-Sbottom-Stop-R model 
with $m_{\sbot}$ and $\mlstop$ close together. In the variants of the $\sbot$ NLSP 
scenario with a heavier $\lstop$ the mode $\lstop \ra b \chonepm$ may open up.

There is no dedicated search for the $\sbot$ NLSP production from gluino decays as yet. We
obtain new constraints on the $\gl -\sbot$ sector in all scenarios  using
the existing  LHC searches involving somewhat similar final states. We 
find that for $m_{sbot}$ around 500 GeV 
the limits on $\mgl$ in the $0l + jets (3b) +\met$ \cite{atlas3b} channel vary in the range 
1.1 - 1.2 TeV in different scenarios (see Fig. \ref{fig2}, \ref{fig3} and Table \ref{tab3}). 
For higher $m_{sbot}$ the limit is degraded rapidly.
On the other hand, for small $\Delta m_{\sbot}$ and $\mgl \approx 1 TeV$, much stronger 
limits on the $\sbot$ mass is obtained via $\gl\gl$ pair production compared to that 
 obtained from  direct $\sbot$ searches. We find that 
the   $0l + jets  +\met$ data generally yield somewhat weaker bounds. 
However, if the relic density  constraint is relaxed and $\Delta m_{\sbot}$ 
is allowed to be smaller ($\approx$ 10 GeV) the strongest limit 
($\approx 1$ TeV)is obtained from the $0l + jets  +\met$ data. 
On the whole the gluino mass limit in the light sbottom scenario 
is about 1 TeV irrespective of $\Delta m_{\sbot}$.

{ \bf Acknowledgments : } 
The work of A. Choudhury, A. Chatterjee and BM was partially supported by 
funding available from the Department of Atomic Energy, Government of India, 
for the Regional Centre for Accelerator-based Particle Physics (RECAPP), 
Harish-Chandra Research Institute. 
AD acknowledges the award of a Senior Scientist position by the 
Indian National Science Academy. AD also thanks RECAPP, HRI for the 
hospitality during the initial phase of the project. 

\end{document}